\documentstyle[psfig,12pt,twoside,fleqn,espcrc1]{article}

\input epsf

\def\ms{$M_\odot$}
\def\ni{$^{56}$Ni}

\def\etal{et al. }

\def\ltsima{$\; \buildrel < \over \sim \;$}
\def\ltsim{\lower.5ex\hbox{\ltsima}}
\def\gtsima{$\; \buildrel > \over \sim \;$}
\def\gtsim{\lower.5ex\hbox{\gtsima}}

\title{\bf NUCLEOSYNTHESIS IN TYPE II SUPERNOVAE}

\author{
K. Nomoto\address{Department of Astronomy, University of Tokyo,Tokyo
 113, Japan},
M. Hashimoto\address{Department of Physics, Faculty of Science,
 Kyushu University, Fukuoka 810, Japan},
T. Tsujimoto\address{National Astronomical Observatory, Mitaka, 
Tokyo 181, Japan},
F.-K. Thielemann\address{Institut f\"ur
Theoretische Physik, Universit\"at Basel CH-4056 Basel, Switzerland},
N. Kishimoto{\hbox{$^{\rm a}$}},\\
Y. Kubo{\hbox{$^{\rm a}$}}, N. Nakasato{\hbox{$^{\rm a}$}}
}

\begin{document}
\maketitle

\begin{abstract}

     Presupernova evolution and explosive nucleosynthesis in massive   
stars for main-sequence masses from 13 $M_\odot$ to 70   
$M_\odot$ are calculated.  We examine the dependence of the  supernova  
yields on the stellar mass, ${\rm  ^{12}C(\alpha, \gamma) ^{16}O}$ 
rate, and explosion energy.  The supernova yields integrated over the  
initial mass function are compared with the solar abundances.  

\end{abstract}

\section {Stellar Nucleosynthesis and the ${\rm ^{12}C(\alpha, 
\gamma)^{16}O}$ Rate}

     Nucleosynthesis in massive stars is one of the major sources of
nuclei in the cosmos.  We present presupernova models for
helium stars with masses of $M_\alpha$ = 3.3, 4, 5, 6, 8, 16, and 32
$M_\odot$ as an extension of the studies by Nomoto \& Hashimoto
(1988).  These helium star masses correspond approximately to
main-sequence masses of $M_{\rm ms}$ = 13, 15, 18, 20, 25, 40, and 70
$M_\odot$, respectively (Sugimoto \& Nomoto 1980).  The systematic
study for such a dense grid of stellar masses enables us to understand
how explosive nucleosynthesis depends on the presupernova stellar
structure and to apply the results to the chemical evolution of
galaxies.  We use the Schwarzschild criterion for convection and
neglect overshooting.  The initial compositions are: $X({\rm ^4He})$ =
0.9879 and $X({\rm ^{14}N})$ = 0.0121, where all the original CNO
elements are assumed to be converted into $^{14}$N during core
hydrogen burning.  These helium stars are evolved from helium burning
through the onset of the Fe core collapse.

     Nuclear reaction rates are mostly taken from Caughlan \& Fowler  
(1988).  For the uncertain rate of ${\rm ^{12}C(\alpha,  
\gamma)^{16}O}$, we use the rate by Caughlan \etal (1985; CFHZ85),  
which is larger than the rate by Caughlan \& Fowler (1988; CF88) by a  
factor of $\sim 2.3$.  To examine the influence of this difference, we  
evolve the $M_\alpha$ = 8 \ms\ helium star, using the ${\rm  
^{12}C(\alpha, \gamma)^{16}O}$ rate by CF88 (case 25B).  [The 25 \ms\  
star model with the ${\rm ^{12}C(\alpha, \gamma)^{16}O}$ rate by CFHZ85  
is denoted as case 25A.]  At the end of core helium burning, the  
formation of the carbon-oxygen core and its composition are influenced  
largely by the ${\rm ^{12}C(\alpha, \gamma)^{16}O}$ rate.  The larger rate  
results in a smaller C/O ratio, which affects the abundances of Ne,  
Mg, Al relative to O in the more evolved cores.  

     Comparison of the presupernova density structures for the two
cases 25A and 25B shows that case 25B has a more concentrated core at
$M_r < 2 M_\odot$ (i.e., a steeper density gradient) and more extended
outer layers than case 25A.  This is due to a
larger carbon abundance and thus stronger carbon shell burning for
25B.  Model 25B has smaller
masses of the Fe core (1.37 \ms) and the O-rich layer than those for
25A (1.41 \ms) due also to the stronger carbon shell burning.

     It is found that the size of the iron core is not a monotonic
function of the helium core mass as shown by Barkat \& Marom (1990)
and Woosley \& Weaver (1995).  For $M_{\rm ms}$ = 13, 15, 18, 20, 25
(case 25A), 40, and 70 $M_\odot$, the iron core masses are 1.18, 1.28,
1.36, 1.40, 1.42, 1.88, and 1.57 $M_\odot$, respectively.  In case
25B, the iron core mass is 1.37 $M_\odot$, which is smaller than in
case 25A.

\section {Explosive Nucleosynthesis}
 
     The hydrodynamic phases of supernova explosions for the above 
eight presupernova models were followed with an extensive nuclear 
reaction network (Hashimoto \etal 1989, 1993; Thielemann \etal 1990, 
1996).
 
     Since the mechanism of supernova explosions after core collapse is 
not fully understood yet, the explosion energy and the mass cut (or 
\ni\ mass) have remaining uncertainties, except for SN 1987A.  The 
final kinetic energy of the explosion is assumed to be $E$ = 1.0 
$\times$ 10$^{51}$ erg as inferred from the modeling of SN 1987A and SN 
1993J (e.g., Shigeyama \& Nomoto 1990; Shigeyama \etal 1994).  

     In the present study, the mass cut is chosen to produce $\sim$
0.15 $M_\odot$ ${\rm ^{56}Ni}$ for 13 -- 15 \ms\ stars and $\sim$
0.075 $M_\odot$ ${\rm ^{56}Ni}$ for 18 -- 70 \ms\ stars.  This is
based on the estimates from the light curves of SN 1993J and Type Ib
supernovae for the 13 -- 15 $M_\odot$ stars (e.g., Nomoto \etal 1993)
and SN 1987A for the 18 -- 20 $M_\odot$ stars (e.g., Nomoto \etal
1993).  For more massive stars, a similar mass of ${\rm ^{56}Ni}$ is
suggested from SN 1990E (Schmidt \etal 1993).  
 
\begin{figure}[p]
\hspace*{1.5cm}
\epsfxsize=300pt
\epsfysize=200pt
\epsfbox{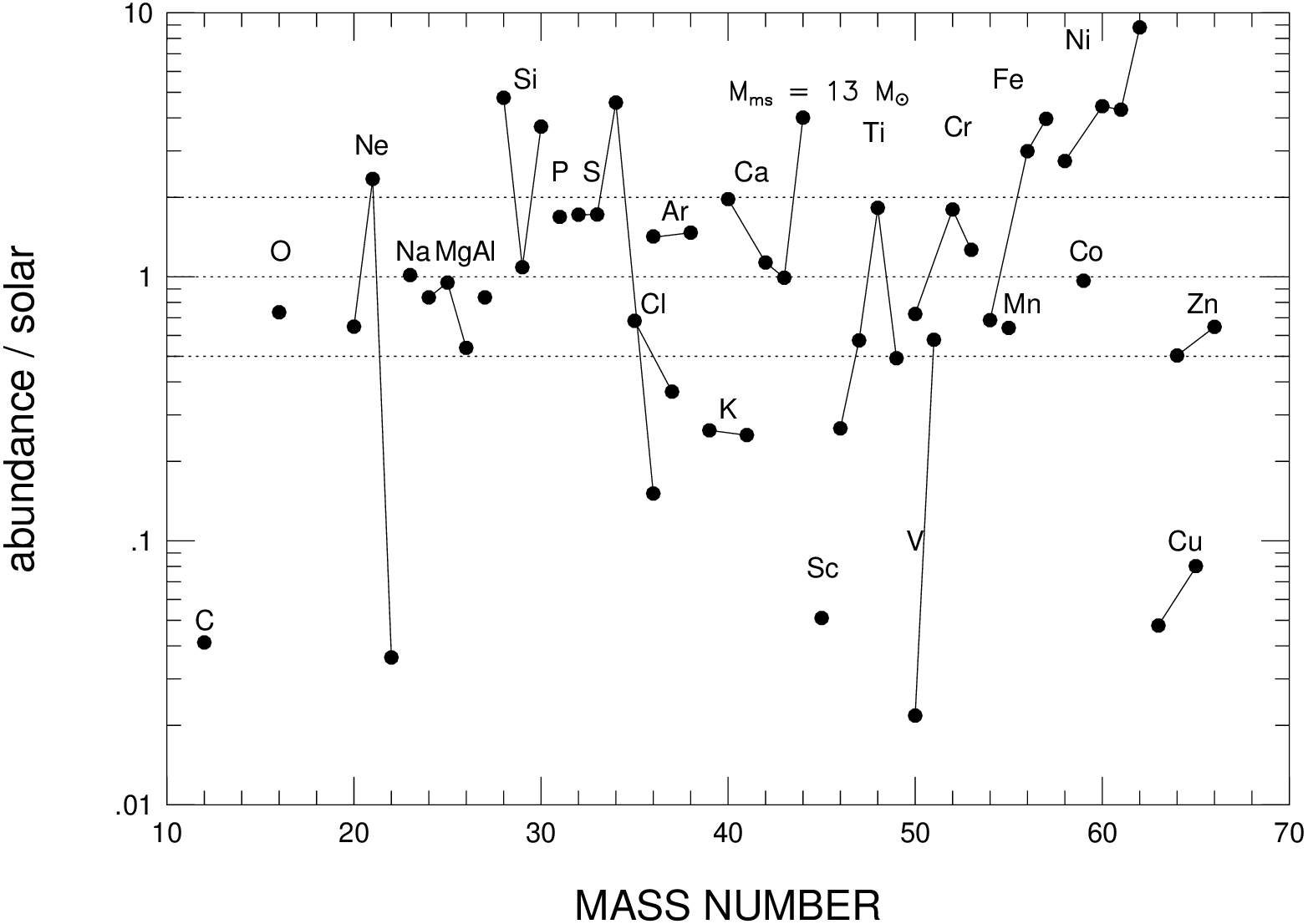}
\vspace{4mm} \\
\hspace*{1.5cm}
\epsfxsize=300pt
\epsfysize=200pt
\epsfbox{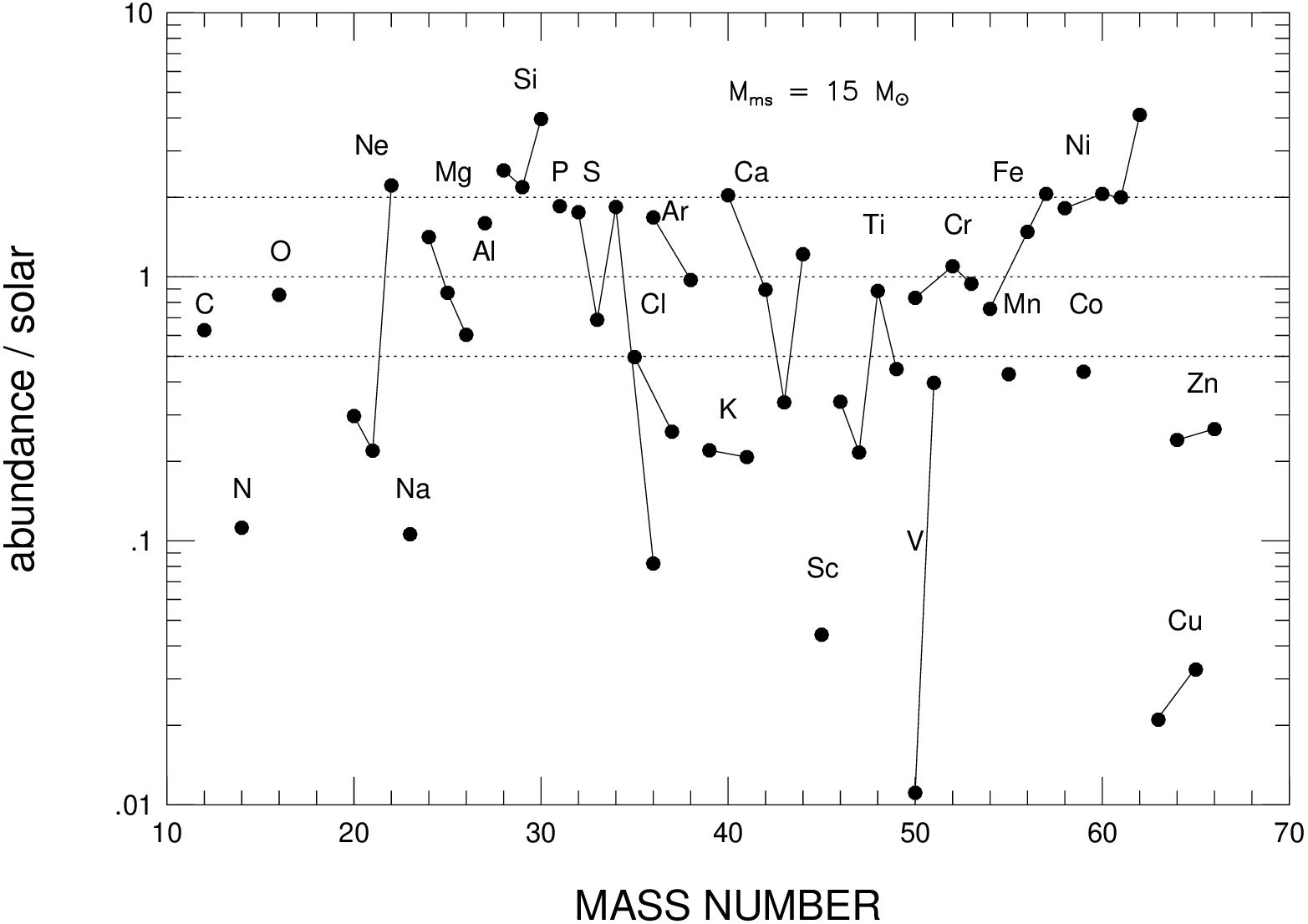}
\vspace{4mm} \\
\hspace*{1.5cm}
\epsfxsize=300pt
\epsfysize=200pt
\epsfbox{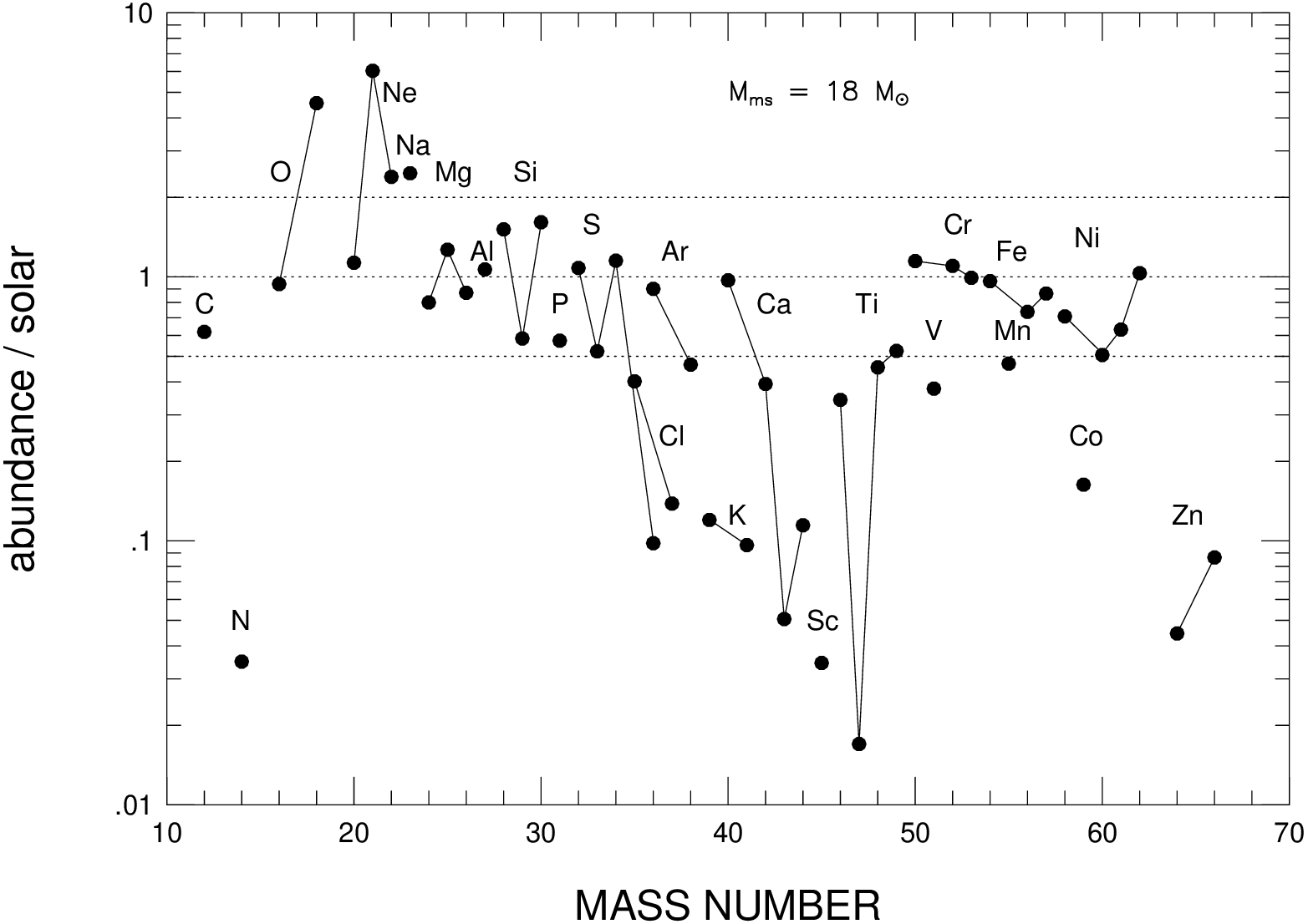}                                              
\caption{
Abundances of stable isotopes relative to the solar values 
 for the 13, 15, and 18 \ms\ stars (H-rich envelope is not 
included).
} 
\label{fig:abd1}
\end{figure}

\begin{figure}[p]
\hspace*{1.5cm}
\epsfxsize=300pt
\epsfysize=200pt
\epsfbox{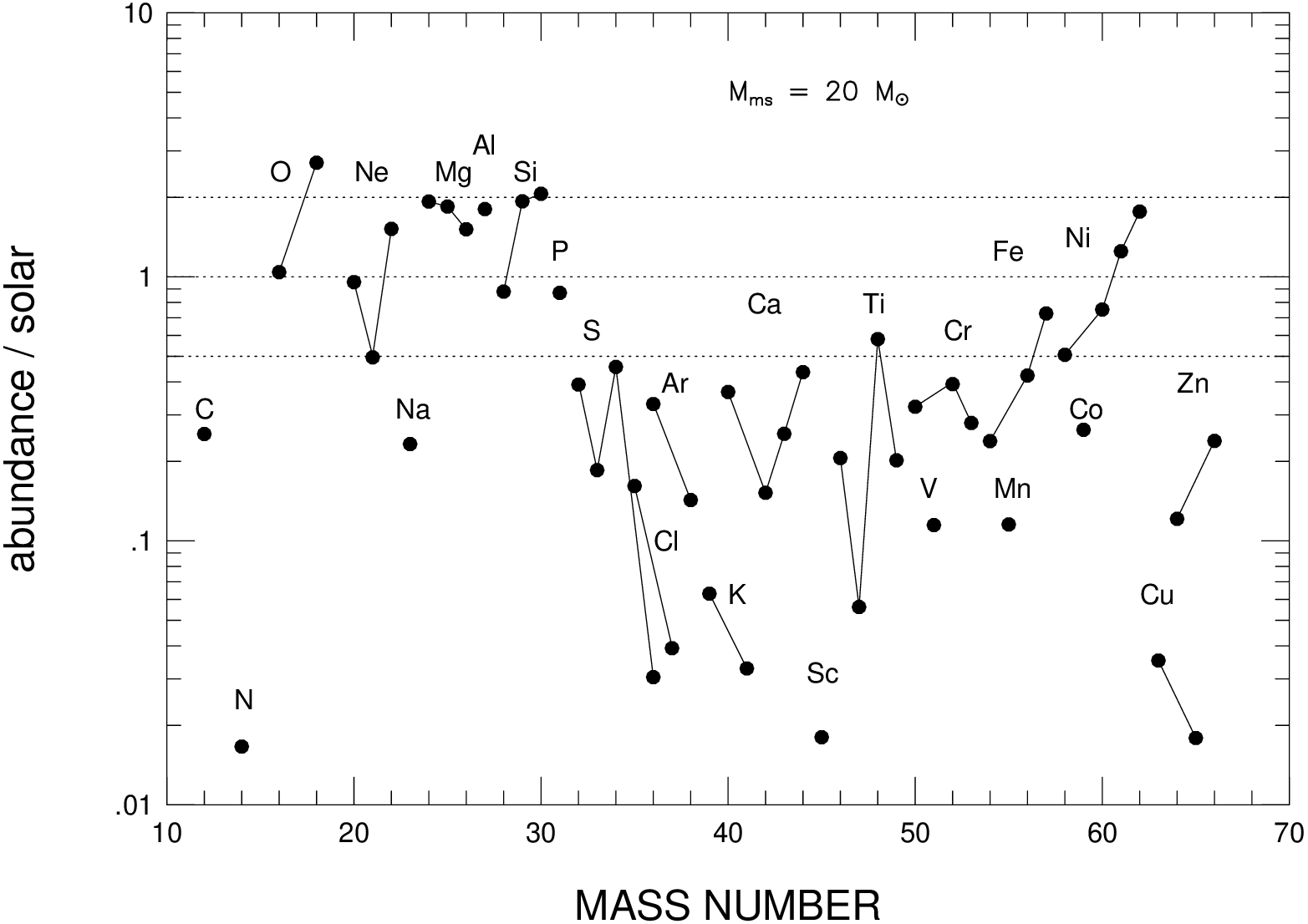}
\vspace{4mm} \\
\hspace*{1.5cm}
\epsfxsize=300pt
\epsfysize=200pt
\epsfbox{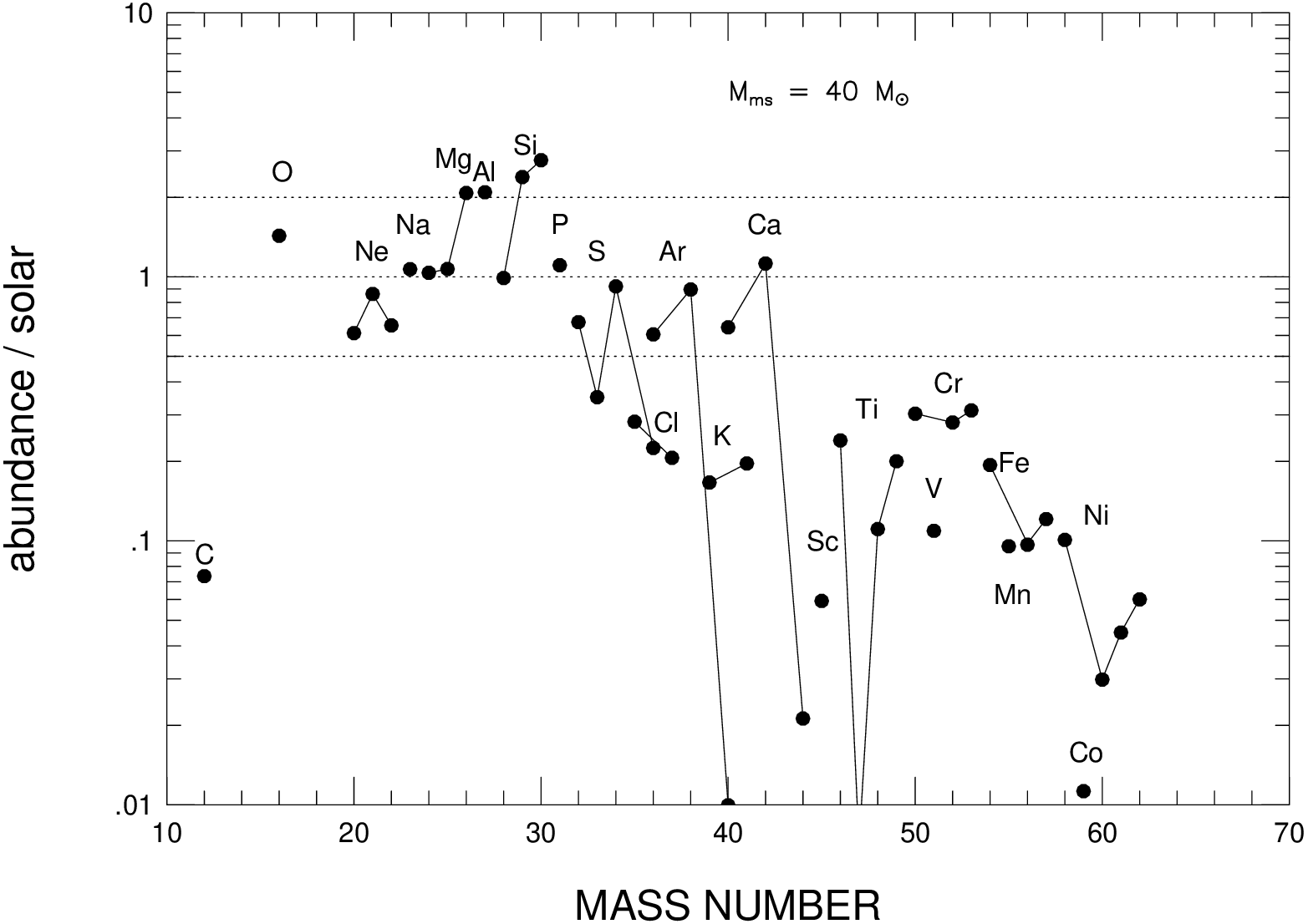}
\vspace{4mm} \\
\hspace*{1.5cm}
\epsfxsize=300pt
\epsfysize=200pt
\epsfbox{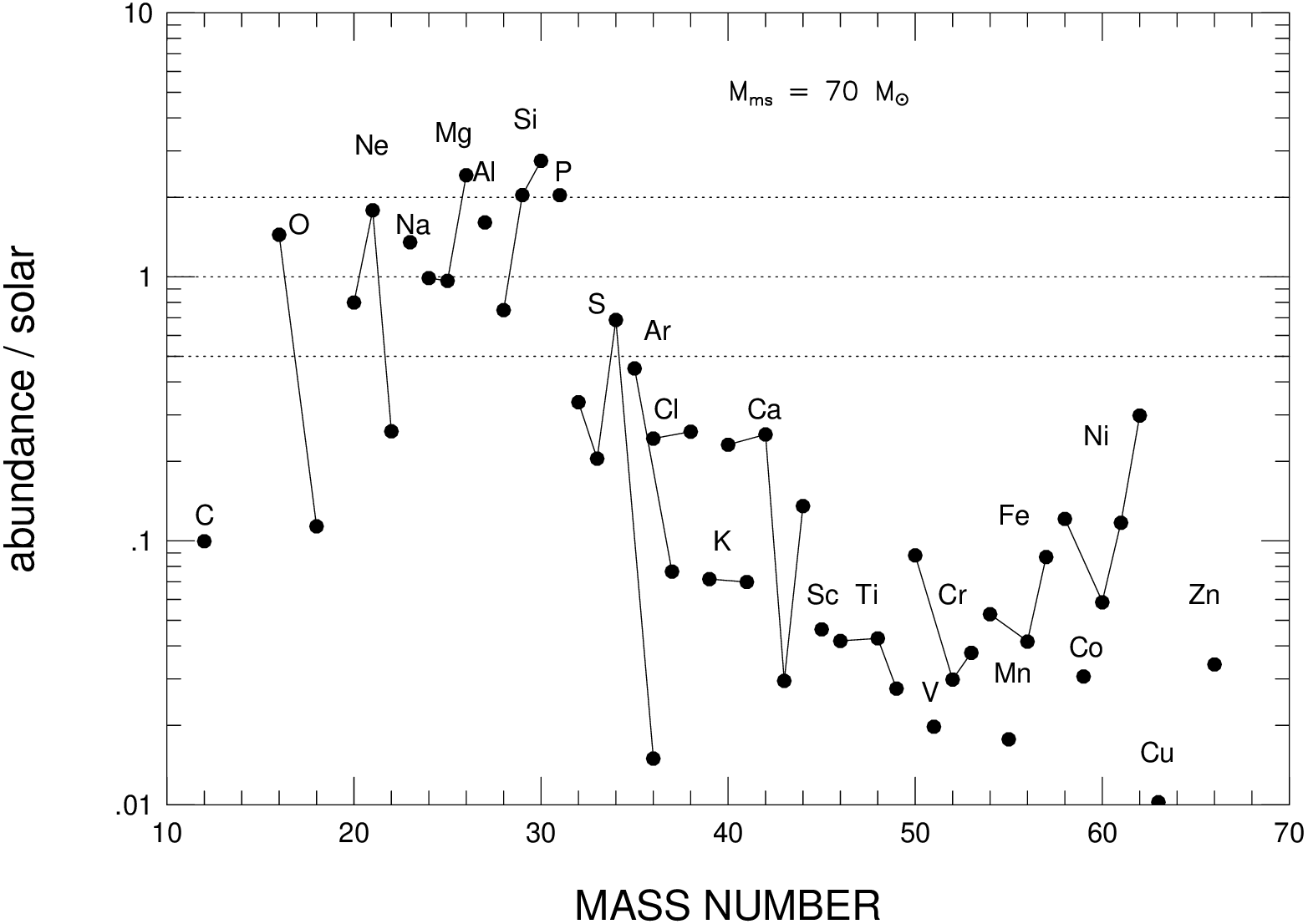}
\caption{
Same as Fig. \ref{fig:abd1} but for the 20, 40, and 70 \ms\ stars.
}
\label{fig:abd2}
\end{figure}

     Figures \ref{fig:abd1} and \ref{fig:abd2} show the
integrated abundances of the ejecta relative to the solar values
(Anders \& Grevesse 1989) for $M_{\rm ms}$ = 13, 15, 18, 20, 40, and
70 $M_\odot$.  Figure \ref{fig:abd3} shows three cases of $M_{\rm ms}$
= 25 $M_\odot$, i.e., cases 25A, 25B, and 25BE (see below).  Table 1
gives the ejected masses (\ms) of stable species for the 13 - 70 \ms\
stars.

     To examine the dependence on the explosion energy, we show the 
case 25BE, i.e., case 25B with $E = 1.5 \times 10^{51}$ erg.  The  
larger explosion energy leads to the outward shift of the abundance 
distribution.  This leads to minor differences between the abundances 
for the two explosion energies (Fig. \ref{fig:abd3}).

\section {Isotopic Abundances}
 
\begin{figure}[t]
\hspace*{1.5cm}
\epsfxsize=300pt
\epsfysize=200pt
\epsfbox{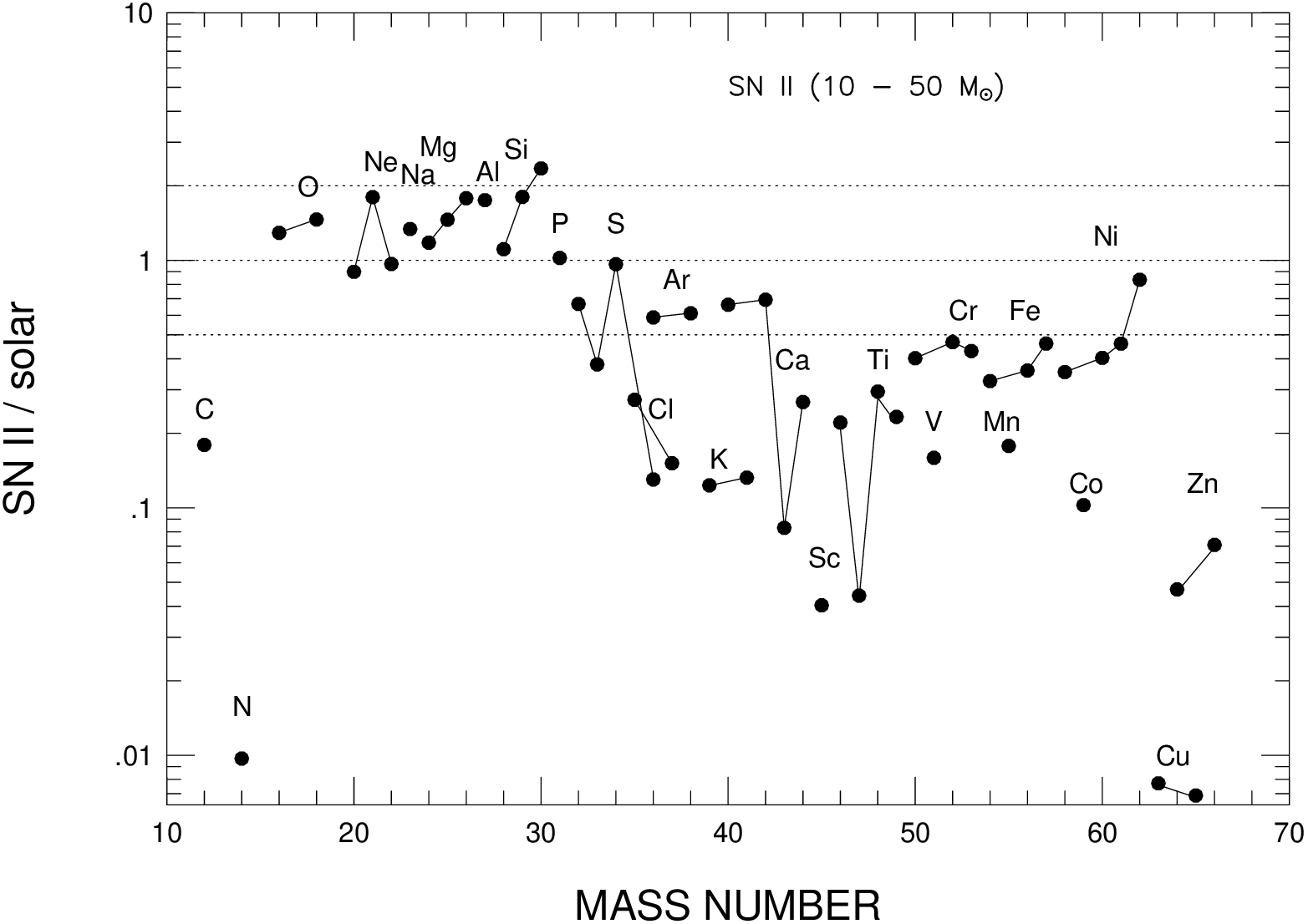}
\caption{
Nucleosynthesis products from 10 -- 50 \ms\ stars averaged over the IMF 
relative to solar abundances.
}
\label{fig:abd4}
\end{figure}

     Figure \ref{fig:abd4} shows the isotopic abundances relative to  
their solar values (Anders \& Grevesse 1989) after averaging over the  
mass range from 10 to 50 $M_\odot$ with an initial mass function  
$\propto M^{-1.35}$.  Here the upper mass limit 50 \ms\ is chosen from 
the comparison of [O/Fe] and [Mg/Fe] with those of metal-poor stars 
(Tsujimoto \etal 1993).  We also assume no heavy element production 
below 10 \ms\ and approximate the abundances of 10 -- 13 \ms\ stars by 
a linear interpolation between 10 and 13 \ms.  

     Figure \ref{fig:abd4} shows that the relative abundance ratios 
>from massive stars are in good agreement with the solar ratios for $A < 
27$.  [The sum of type Ia and type II products with a ratio of 1 to 9  
reproduces well the solar abundances for a wider range of $A$ 
(Tsujimoto \etal 1995).]  Note that this agreement is realized for the 
${\rm  ^{12}C(\alpha, \gamma) ^{16}O}$ rate by CFHZ85, i.e., case 25A.   

     For case 25B, Ne, Na, and Al relative to O are overproduced with 
respect to the solar ratios as seen in Figure \ref{fig:abd3}.  This is 
due to the larger C/O ratio in case 25B after helium burning.  Since  
the products of the 25 $M_\odot$ star dominate type II supernova 
yields, this result suggests that the ${\rm ^{12}C(\alpha, 
\gamma)^{16}O}$ rate is higher than that of CF88 and closer to CFHZ85.  
The presently most reliable experimental investigations give values 
inbetween the two rates.

     We should note that the isotopic ratios in Figures \ref{fig:abd1}  
-- \ref{fig:abd4} depend not only on the ${\rm ^{12}C(\alpha,  
\gamma)^{16}O}$ rate but also on convective overshooting, mixing fresh 
He into the core at late high temperature core helium burning stages.  
The above  comparison that favors the CFHZ85 rate is based on the 
calculations with no convective overshooting.  If  overshooting during 
convective core helium burning would reduce the C/O  ratio, a smaller 
${\rm ^{12}C(\alpha, \gamma)^{16}O}$ rate would be favored (Weaver \& 
Woosley 1993; Woosley \& Weaver 1995).

     Figure \ref{fig:abd4} also shows that some species, ${\rm
^{35}Cl}$, ${\rm ^{39}K}$, ${\rm ^{44}Ca}$, ${\rm ^{48}Ti}$, and ${\rm
^{59}Co}$, are underproduced relative to the solar values.  If we
include the weak component of the s-process nuclei $50 < A < 100$
produced during core helium burning (Prantzos \etal 1990), ${\rm
^{48}Ti}$ and ${\rm ^{59}Co}$ are enhanced appreciably compared with
the seed (solar) abundances.  ${\rm ^{35}Cl}$, ${\rm ^{39}K}$ and
${\rm ^{44}Ca}$ are enhanced only by a factor of $\sim$ 2.  Synthesis
of s-process elements during carbon shell burning would also be  
significant.

\section*{Acknowledgements}

This work has been supported in part by the Grant in Aid for
Scientific Research (05243206, 06234210) and COE research (07CE2002)
of the Ministry of Education, Science, and Culture in Japan.

\begin{figure}[p]
\hspace*{1.5cm}
\epsfxsize=300pt
\epsfysize=200pt
\epsfbox{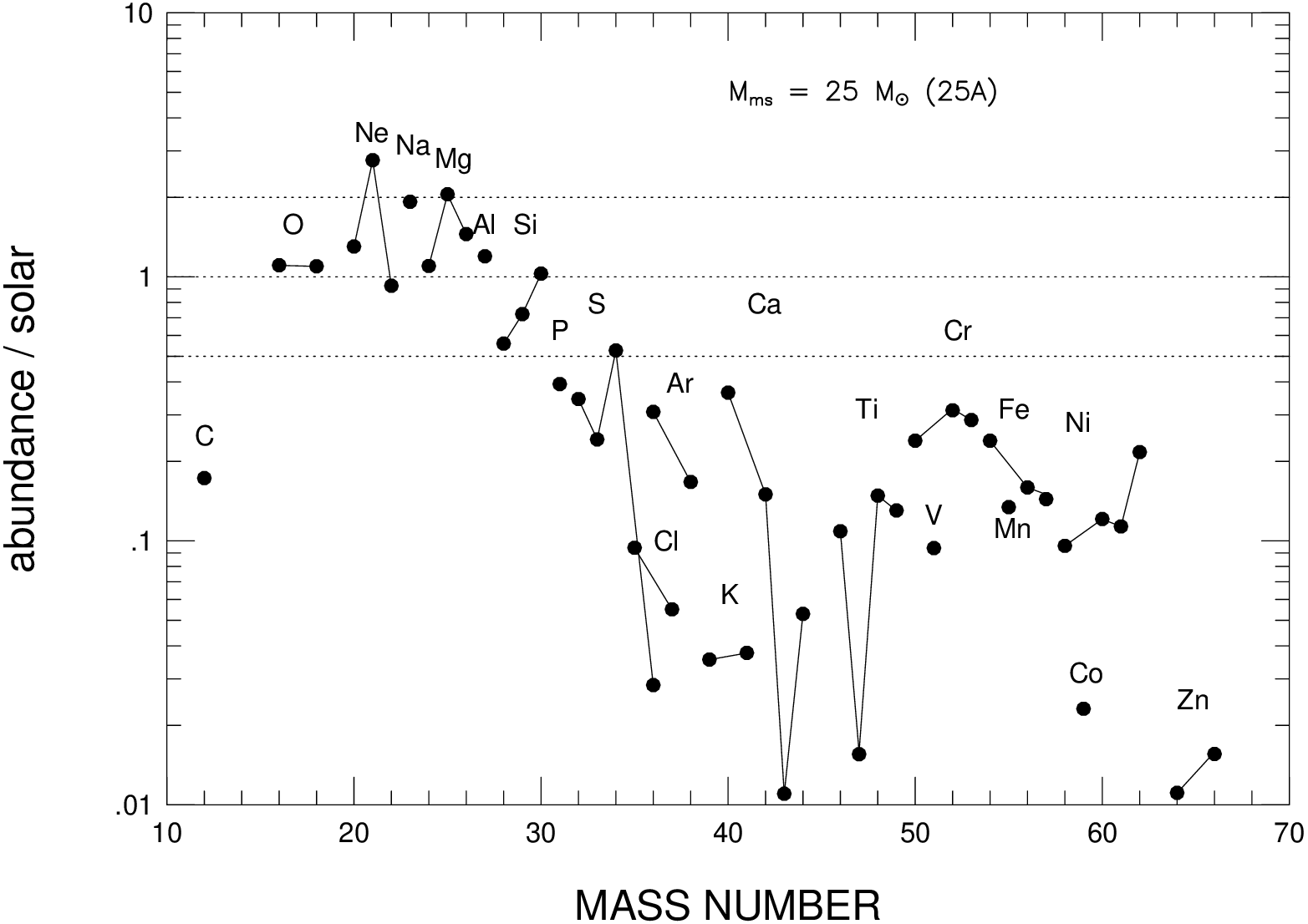}
\vspace{4mm} \\
\hspace*{1.5cm}
\epsfxsize=300pt
\epsfysize=200pt
\epsfbox{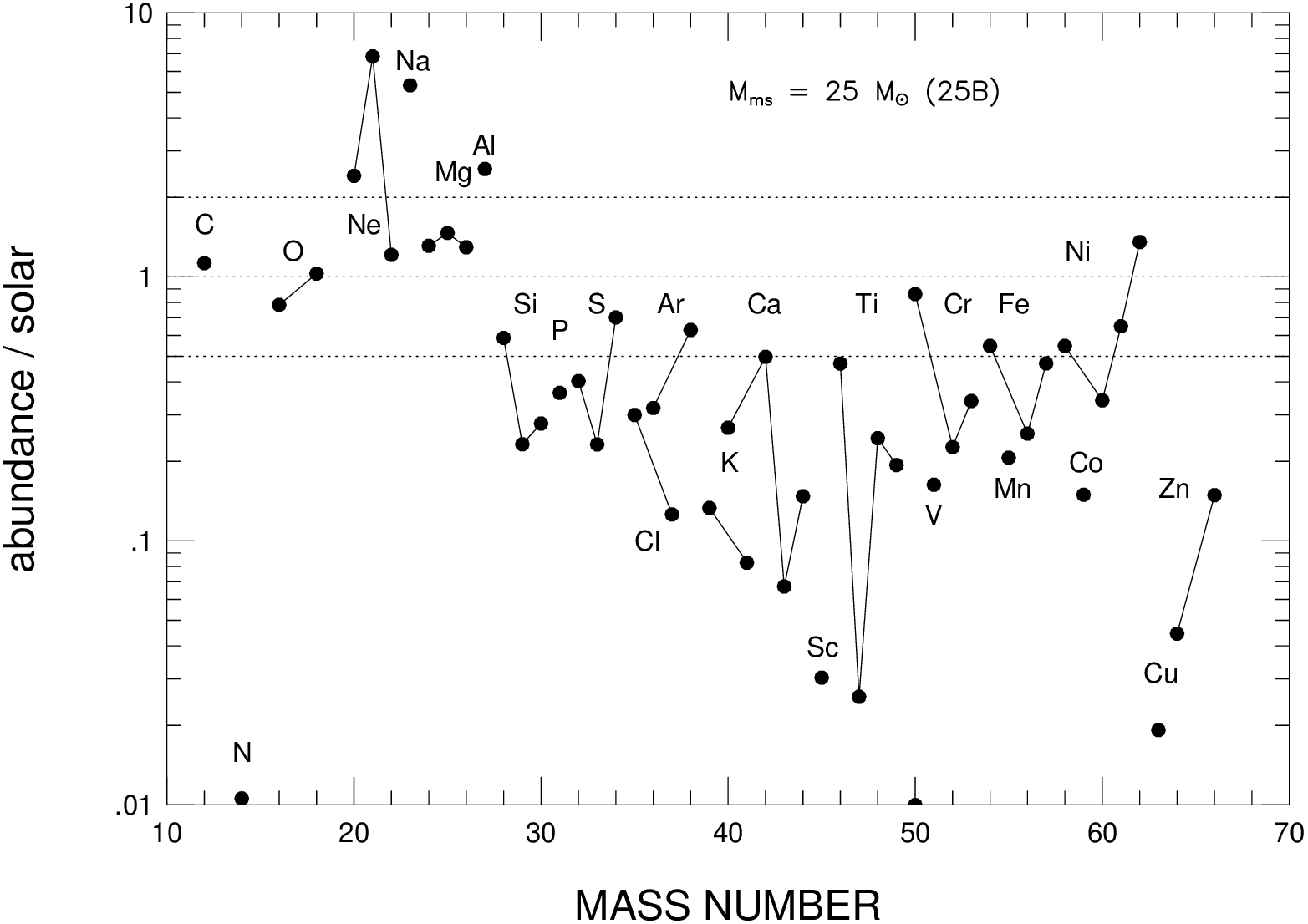}
\vspace{4mm} \\
\hspace*{1.5cm}
\epsfxsize=300pt
\epsfysize=200pt
\epsfbox{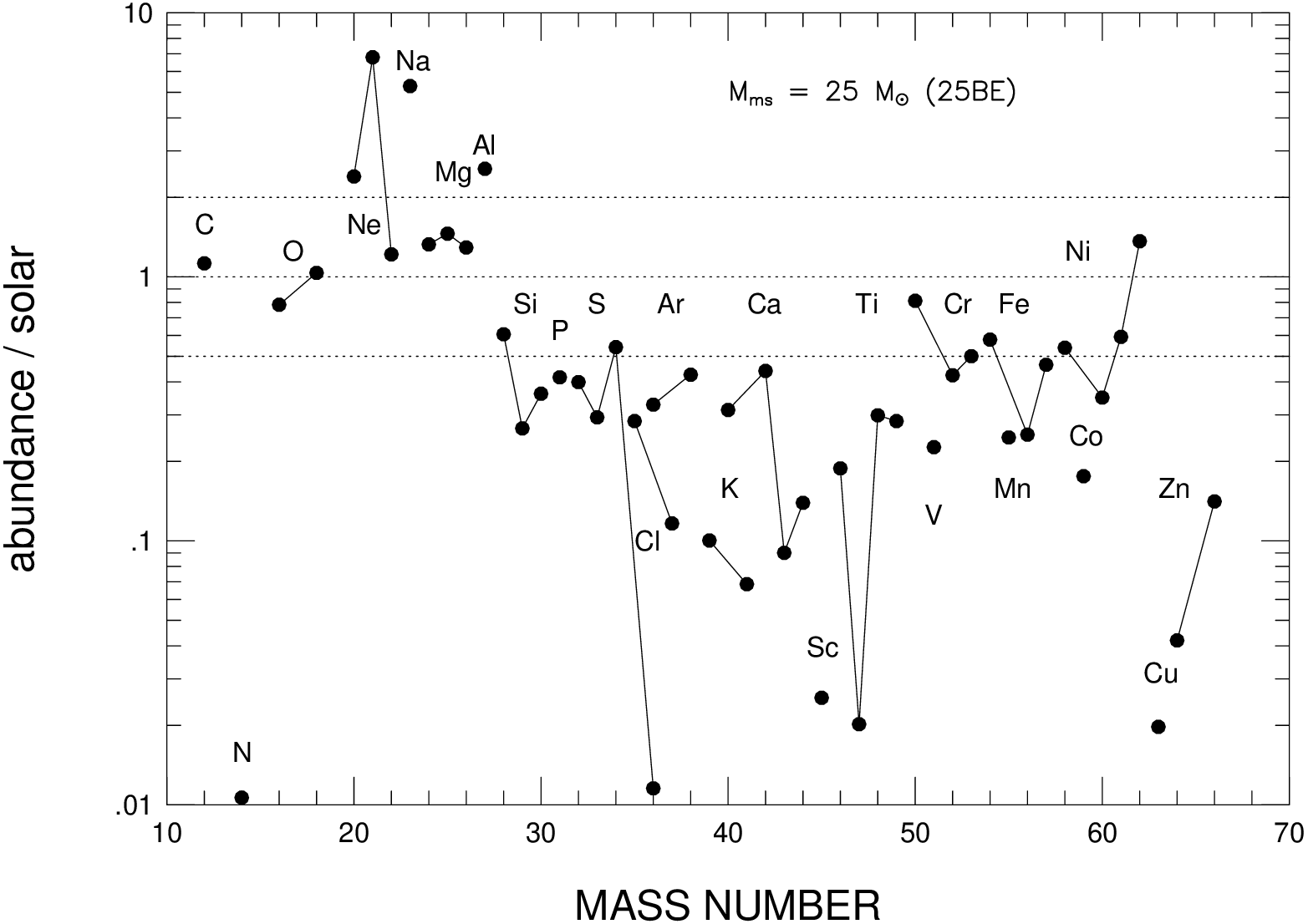}
\caption{
Same as in Fig. \ref{fig:abd1} but for the 25 \ms\ star (case 25A).   
Case 25B uses the ${\rm ^{12}C(\alpha, \gamma)^{16}O}$ rate by CF88 
and case 25BE is the same as 25B but with $E$ = 1.5 $\times$ $10^{51}$  
erg s$^{-1}$.}
\label{fig:abd3}
\end{figure}
\vfil\eject


\begin{table}[h]
\caption{Nucleosynthesis products of SNe II for various progenitor 
masses (H-rich envelope is not included).
}

{\small
\begin{tabular}{llllllll}
\hline \hline
& \multicolumn{7}{c}{Synthesized isotopic mass ($M_\odot$)} \\
\cline{2-8}
Species & $m$=13 $M_\odot$ & $m$=15 $M_\odot$ & $m$=18 $M_\odot$ 
& $m$=20 $M_\odot$ & 
$m$=25 $M_\odot$ & $m$=40 $M_\odot$ & $m$=70 $M_\odot$ \\ \hline
$^{12}$C & 2.68E-03 & 8.26E-02 & 1.65E-01 & 1.14E-01 & 1.48E-01 & 
1.48E-01 & 4.67E-01 \\
$^{13}$C & 9.47E-09 & 4.97E-10 & 7.73E-10 & 1.17E-10 & 1.03E-08 &
3.02E-10 & 2.57E-10 \\ 
$^{14}$N & 3.75E-08 & 5.37E-03 & 3.39E-03 & 2.72E-03 & 9.53E-04 &
7.08E-05 & 7.68E-03 \\
$^{15}$N & 2.08E-08 & 1.36E-10 & 9.05E-08 & 6.48E-10 & 1.04E-08 &
1.19E-08 & 2.36E-10 \\
$^{16}$O & 1.51E-01 & 3.55E-01 & 7.92E-01 & 1.48 & 2.99 & 9.11 & 
2.14E+01 \\
$^{17}$O & 6.07E-08 & 4.41E-09 & 4.01E-07 & 9.86E-09 & 7.86E-08 &
3.13E-07 & 6.64E-10 \\
$^{18}$O & 9.44E-09 &  1.35E-02 & 8.67E-03 & 8.68E-03 & 6.69E-03 & 
1.79E-06 & 3.80E-03 \\ 
$^{19}$F & 8.06E-10 & 2.12E-11 & 7.67E-09 & 7.84E-11 & 8.17E-10 &
7.38E-10 & 2.63E-15 \\
$^{20}$Ne & 2.25E-02 & 2.08E-02 & 1.61E-01 & 2.29E-01 & 5.94E-01 & 
6.58E-01 & 2.00 \\
$^{21}$Ne & 2.08E-04 & 3.93E-05 & 2.19E-03 & 3.03E-04 & 3.22E-03 & 
2.36E-03 & 1.14E-02 \\
$^{22}$Ne & 1.01E-04 & 1.25E-02 & 2.74E-02 & 2.93E-02 & 3.39E-02 & 
5.66E-02 & 5.23E-02 \\ 
$^{23}$Na & 7.27E-04 & 1.53E-04 & 7.25E-03 & 1.15E-03 & 1.81E-02 & 
2.37E-02 & 6.98E-02 \\ 
$^{24}$Mg & 9.23E-03 & 3.16E-02 & 3.62E-02 & 1.47E-01 & 1.59E-01 & 
3.54E-01 & 7.87E-01 \\
$^{25}$Mg & 1.38E-03 & 2.55E-03 & 7.54E-03 & 1.85E-02 & 3.92E-02 & 
4.81E-02 & 1.01E-01 \\ 
$^{26}$Mg & 8.96E-04 & 2.03E-03 & 5.94E-03 & 1.74E-02 & 3.17E-02 & 
1.07E-01 & 2.91E-01 \\
$^{27}$Al & 1.04E-03 & 4.01E-03 & 5.44E-03 & 1.55E-02 & 1.95E-02 & 
8.05E-02 & 1.44E-01 \\ 
$^{28}$Si & 6.68E-02 & 7.16E-02 & 8.69E-02 & 8.50E-02 & 1.03E-01 & 
4.29E-01 & 7.55E-01 \\
$^{29}$Si & 7.99E-04 & 3.25E-03 & 1.76E-03 & 9.80E-03 & 6.97E-03 & 
5.43E-02 & 1.08E-01 \\  
$^{30}$Si & 1.87E-03 & 4.04E-03 & 3.33E-03 & 7.19E-03 & 6.81E-03 & 
4.32E-02 & 1.00E-01 \\
$^{31}$P & 2.95E-04 & 6.55E-04 & 4.11E-04 & 1.05E-03 & 9.02E-04 & 
5.99E-03 & 2.57E-02 \\ 
\hline
\end{tabular}
}
\end{table}

\begin{table}[h]
{\small
\begin{tabular}{llllllll}
\hline \hline
Species & $m$=13 $M_\odot$ & $m$=15 $M_\odot$ & $m$=18 $M_\odot$ 
& $m$=20 $M_\odot$ & 
$m$=25 $M_\odot$ & $m$=40 $M_\odot$ & $m$=70 $M_\odot$ \\ \hline
$^{32}$S & 1.46E-02 & 3.01E-02 & 3.76E-02 & 2.29E-02 & 3.84E-02 & 
1.77E-01 & 2.05E-01 \\
$^{33}$S & 1.19E-04 & 9.60E-05 & 1.48E-04 & 8.84E-05 & 2.20E-04 & 
7.49E-04 & 1.02E-03 \\
$^{34}$S & 1.83E-03 & 1.49E-03 & 1.89E-03 & 1.26E-03 & 2.77E-03 & 
1.14E-02 & 1.98E-02 \\
$^{36}$S & 3.04E-07 & 3.34E-07 & 8.08E-07 & 4.23E-07 & 7.51E-07 &
1.40E-05 & 2.17E-06 \\
$^{35}$Cl & 3.70E-05 & 3.45E-05 & 8.95E-05 & 
6.05E-05 & 6.72E-05 & 4.75E-04 & 1.76E-03 \\  
$^{37}$Cl & 6.73E-06 & 9.60E-06 & 1.04E-05 &  
4.96E-06 & 1.32E-05 & 1.17E-04 & 1.01E-04 \\  
$^{36}$Ar & 2.36E-03 & 5.63E-03 & 6.13E-03 & 3.78E-03 & 6.71E-03 & 
3.11E-02 & 2.92E-02 \\
$^{38}$Ar & 4.85E-04 & 6.49E-04 & 6.29E-04 & 3.25E-04 & 7.24E-04 & 
9.14E-03 & 6.16E-03 \\ 
$^{40}$Ar & 4.82E-09 & 3.24E-09 & 1.42E-08 & 4.65E-09 & 8.92E-09 &
1.74E-07 & 5.07E-08 \\
$^{39}$K & 1.95E-05 & 3.31E-05 & 3.66E-05 & 
3.24E-05 & 3.47E-05 & 3.83E-04 & 3.84E-04 \\
$^{41}$K & 1.42E-06  & 2.37E-06  & 2.23E-06  & 
1.28E-06  & 2.79E-06 & 3.43E-05  & 2.84E-05  \\
$^{40}$Ca & 2.53E-03 & 5.29E-03 & 5.11E-03 & 3.25E-03 & 6.15E-03 & 
2.56E-02 & 2.14E-02 \\
$^{42}$Ca & 1.02E-05 & 1.63E-05 & 1.45E-05 & 9.45E-06 & 1.77E-05 &
3.13E-04 & 1.64E-04 \\
$^{43}$Ca & 1.91E-06 & 1.30E-06 & 3.99E-07 & 3.38E-06 & 2.78E-07 &
4.02E-07 & 4.09E-06 \\
$^{44}$Ca & 1.22E-04  & 7.49E-05  & 1.43E-05  & 
9.15E-05  & 2.11E-05  & 2.00E-05  & 2.97E-04  \\
$^{46}$Ca & 2.06E-10 & 6.23E-11 & 3.23E-11 & 1.12E-11 & 2.60E-10 &
4.39E-10 & 2.23E-10 \\
$^{48}$Ca & 1.13E-13 & 3.99E-16 & 1.07E-15 & 2.41E-16 & 1.70E-14 &
2.48E-13 & 2.36E-14 \\
$^{45}$Sc & 4.26E-08 & 7.44E-08 & 1.18E-07 & 1.04E-07 & 8.96E-08 &
1.53E-06 & 2.78E-06 \\
$^{46}$Ti & 2.56E-06  & 6.26E-06  & 6.72E-06  & 
6.81E-06  & 6.84E-06  & 3.56E-05  & 1.44E-05  \\ 
$^{47}$Ti & 5.13E-06  & 3.75E-06  & 3.11E-07  & 
1.73E-06  & 9.11E-07  & 9.74E-07  & 6.26E-07  \\ 
$^{48}$Ti & 1.68E-04  & 1.58E-04  & 8.59E-05  & 
1.85E-04  & 8.98E-05  & 1.58E-04  & 1.42E-04  \\ 
$^{49}$Ti & 3.45E-06  & 6.10E-06  & 7.54E-06  & 
4.89E-06  & 6.01E-06  & 2.17E-05  & 6.97E-06  \\ 
$^{50}$Ti & 3.56E-10  & 1.21E-09  & 1.17E-10  & 
1.12E-10  & 5.90E-10  & 2.00E-10  & 2.56E-10  \\ 
$^{50}$V & 8.65E-10 & 8.57E-10 & 4.64E-10 & 2.15E-10 & 7.99E-10 &
2.14E-09 & 1.52E-09 \\
$^{51}$V & 9.34E-06 & 1.25E-05 & 1.25E-05 & 6.40E-06 & 9.96E-06 &
2.73E-05 & 1.15E-05 \\
$^{50}$Cr & 2.30E-05 & 5.15E-05 & 7.49E-05 & 3.54E-05 & 5.01E-05 & 
1.49E-04 & 1.01E-04 \\
$^{52}$Cr & 1.15E-03 & 1.36E-03 & 1.44E-03 & 8.64E-04 & 1.31E-03 & 
2.77E-03 & 6.86E-04 \\
$^{53}$Cr & 9.34E-05 & 1.35E-04 & 1.50E-04 & 7.12E-05 & 1.39E-04 & 
3.56E-04 & 1.00E-04 \\ 
$^{54}$Cr & 3.35E-08 & 4.09E-08 & 2.53E-08 & 6.26E-09 & 2.41E-08 & 
2.81E-08 & 7.61E-08 \\ 
$^{55}$Mn & 3.65E-04 & 4.74E-04 & 5.48E-04 & 2.27E-04 & 5.02E-04 & 
8.41E-04 & 3.64E-04 \\ 
$^{54}$Fe & 2.10E-03 & 4.49E-03 & 6.04E-03 & 2.52E-03 & 4.81E-03 & 
9.17E-03 & 5.81E-03 \\
$^{56}$Fe & 1.50E-01 & 1.44E-01 & 7.57E-02 & 7.32E-02 & 5.24E-02 & 
7.50E-02 & 7.50E-02 \\
$^{57}$Fe & 4.86E-03 & 4.90E-03 & 2.17E-03 & 3.07E-03 & 1.16E-03 & 
2.29E-03 & 3.83E-03 \\ 
$^{58}$Fe & 3.93E-09 & 1.27E-08 & 1.37E-08 & 3.70E-09 & 8.34E-09 &
1.29E-08 & 4.17E-08 \\
$^{59}$Co & 1.39E-04  & 1.22E-04  & 4.82E-05  & 
1.31E-04  & 2.19E-05  & 2.51E-05  & 1.59E-04  \\ 
$^{58}$Ni & 5.82E-03 & 7.50E-03 & 3.08E-03 & 3.71E-03 & 1.33E-03 & 
3.31E-03 & 9.25E-03 \\
$^{60}$Ni & 3.72E-03 & 3.36E-03 & 8.71E-04 & 2.18E-03 & 6.67E-04 & 
3.88E-04 & 1.77E-03 \\
$^{61}$Ni & 1.58E-04 & 1.43E-04 & 4.77E-05 & 1.59E-04 & 2.75E-05 &
2.57E-05 & 1.55E-04 \\
$^{62}$Ni & 1.05E-03 & 9.50E-04 & 2.52E-04 & 7.26E-04 & 1.70E-04 & 
1.11E-04 & 1.28E-03 \\ 
$^{64}$Ni & 2.02E-15 & 4.28E-15 & 2.93E-16 & 2.06E-15 & 6.08E-15 &
6.49E-16 & 4.33E-12 \\
$^{63}$Cu & 1.18E-06 & 1.01E-06 & 4.32E-07 & 3.00E-06 & 1.50E-07 &
1.62E-07 & 9.09E-06 \\
$^{65}$Cu & 9.11E-07 & 7.17E-07 & 8.40E-08 & 7.02E-07 & 1.42E-07 &
1.89E-08 & 5.34E-07 \\
$^{64}$Zn & 2.14E-05 & 1.99E-05 & 3.89E-06 & 1.78E-05 & 3.10E-06 &
8.79E-07 & 1.02E-05 \\
$^{66}$Zn & 1.63E-05 & 1.30E-05 & 4.47E-06 & 2.08E-05 & 2.58E-06 &
9.99E-07 & 3.09E-05 \\
$^{67}$Zn & 2.13E-08 & 1.54E-08 & 3.39E-09 & 6.39E-08 & 2.95E-09 &
2.51E-10 & 1.95E-07 \\
$^{68}$Zn & 6.63E-09 & 7.35E-09 & 8.36E-10 & 5.33E-09 & 9.29E-10 &
1.20E-10 & 9.51E-08 \\
\hline
\end{tabular}
}

\end{table}

\end{document}